 \documentclass[final,5p,times,twocolumn,numbers]{elsarticle}

\usepackage{amsmath, amssymb}
\usepackage{mathtools, cuted}
\usepackage{lipsum}
\usepackage{graphicx} 
\usepackage{multirow}
\usepackage{fullpage}
\usepackage{tabularx}
\usepackage{booktabs}
\usepackage{hyperref}

\journal{Physics Letters B}

\begin{document}

\begin{frontmatter}



\title{Well-separated soliton-antisoliton pairs with an adjoint Higgs field in 4D space}

\author[first,second]{Liang Gong}

\author[first,second]{Rui Shen\corref{cor1}}
\ead{shen@nju.edu.cn}

\affiliation[first]{organization={National Laboratory of Solid State Microstructures and School of Physics},
            addressline={Nanjing University}, 
            city={Nanjing},
            postcode={210093}, 
            country={China}}

\affiliation[second]{organization={Collaborative Innovation Center of Advanced Microstructures},
            addressline={Nanjing University}, 
            city={Nanjing},
            postcode={210093}, 
            country={China}}
            
\cortext[cor1]{Corresponding author}

\begin{abstract}
We present single soliton states and soliton-antisoliton states with an adjoint Higgs field in 4D flat space. The action of a single soliton state diverges, while the action of soliton-antisoliton states converges. The interaction in a soliton-antisoliton state takes a logarithmic dependence on separation. Such soliton-antisoliton states exhibit stability under a scaling transformation.
\end{abstract}




\end{frontmatter}




\section{Introduction}
\label{sec1}

Soliton-antisoliton interactions are of paramount importance in the realm of topological soliton research \cite{manton}. For instance, magnetic monopoles and antimonopoles in the context of the 3D SU(2) Yang-Mills-Higgs model have garnered significant attention as topological solitons \cite{hooftmonopole, polyakovmonopole}. Possessing spherical symmetry, these solitons exhibit distinctive topological characteristics. In scenarios where the Higgs potential is absent \cite{psmonopole, bogomolnyi}, monopole-antimonopole pairs exhibit a long-range attractive interaction combined with a short-range repulsive interaction \cite{taubes1, taubes2}. The strength of this repulsion is contingent upon the twist angle of the monopole-antimonopole pair. Consequently, a static solution encompassing a monopole and an antimonopole can exist with an appropriate twist angle, often referred to as a ``sphaleron''. The existence of such a state, also considering a nonvanishing scalar potential, has been confirmed through numerical demonstrations \cite{kleihaus}. Moreover, a more intricate numerical exploration of the monopole-antimonopole interactions has been elucidated \cite{saurabh}. Soliton-antisoliton states in a specific model can also be utilized for simulating quark confinement \cite{dvali}. Another example is the superconducting vortex-antivortex interaction, which is primarily characterized by attraction \cite{chaves}. Certain experimental techniques can prevent their annihilation, allowing for the observation of vortex-antivortex states \cite{neal, bobba, ge, prando}. The global vortex-antivortex states in Goldstone model are interesting because while the energy of a single global vortex, also called a string with an additional translation-invariant dimension, diverges logarithmically \cite{PhysRevLett.48.1867}, the vortex-antivortex configuration exhibits finite energy \cite{PhysRevD.47.1324}. The numerical \cite{PhysRevD.47.1324} and analytical \cite{Ovchinnikov_1998} results shows that the interaction potential between the well-separated global vortex and antivortex is also attractive and logarithmically dependent on their separation.

The scaling theory for studying solitons was first proposed by Derrick \cite{derrick} and further clarified by Manton and Sutcliffe \cite{manton}. It is commonly believed that in the 4D Euclidean space, there is no static topological soliton that contains an nontrivial Higgs field and has finite action. Therefore, the 4D soliton with an adjoint Higgs field has not attracted much attention in the study of topological solitons. Instead, researchers' interests are drawn to the instantons in 4D pure Yang-Mills model \cite{manton,rajaraman,schafer}. However, when a soliton and an antisoliton with opposite topological properties appear in pair, the overall configuration remains topologically trivial and may possess finite action. The existence of this configuration needs that the soliton and antisoliton remain stable around their centers, which is not satisfied by the standard 4D SU(2) Yang-Mills-Higgs model. In this paper, we suggest a special 4D model, processing soliton-antisoliton pairs with an adjoint Higgs field. These pairs have finite action, despite the individual soliton or antisoliton exhibiting divergent action.

The letter is organized as follows. We introduce the 4D model and its topological property in Sec.~\ref{sec2}. Spherically symmetric solutions are analyzed in Sec.~\ref{sec3}. In Sec.~\ref{sec4}, we present the ansatz for a well-separated soliton-antisoliton pair, discuss the soliton-antisoliton interaction, and analyse the scaling behavior of the soliton-antisoliton pair. Finally, we briefly conclude in Sec.~\ref{sec5}.

\section{Model and its topological property}
\label{sec2}
The action of our model in 4D Euclidean space with metric tensor $g=\mathrm{diag}(1,1,1,1)$ can be expressed as
    \begin{align}\label{action}
        S_E =& \int d^4 x\bigg[\frac{1}{2}\left(D_{\mu}\Phi\right)^\dagger D^{\mu}\Phi + \lambda\left(|\Phi|^2 - 1\right)^2 \nonumber\\
        &+ \frac{1}{2g^2}\mathrm{Tr}\left(F_{\mu\nu}F^{\mu\nu}\right) \bigg]~,
    \end{align}
where $\lambda$ and $g$ are the parameters of the model. The SU(2) gauge potential is $A_{\mu}=\tau_a A^a_\mu/2$ with $\tau$'s being Pauli matrices and the field strength is
    \begin{equation}\label{gaugestrength}
        F_{\mu\nu} = \partial_\mu A_\nu - \partial_\nu A_\mu + i\left[A_\mu, A_\nu\right]~.
    \end{equation}
The adjoint Higgs field can be written as
    \begin{equation}\label{Higgsfield}
        \Phi = G\begin{pmatrix}1 \\ 0\end{pmatrix} = \begin{pmatrix}\phi^0+i\phi^3 \\ i\phi^1-\phi^2\end{pmatrix}
    \end{equation}
with $G=\phi^0 + i\tau_a\phi^a$ and its covariant derivative is
    \begin{equation}\label{covariantderivative}
        D_\mu\Phi = \partial_\mu\Phi - i A_\mu \Phi~.
    \end{equation}
The minus sign before $A_\mu$ is the only difference from the standard SU(2) Yang-Mills-Higgs model. This could be an effective model after spontaneous symmetry breaking of a complex one, such as the two-Higgs-doublet model~\cite{branco}. We have adopted a specific energy unit such that the vacuum expectation value of the absolute Higgs field $|\Phi|$ is $1$. Taking variations with respect to all fields in order to minimize the action leads to the Euler-Lagrange equations for the fields. $\Phi=(1,0)^T, A_\mu=0$ is the vacuum state with $S_E=0$.

When discussing solitons and antisolitons, their topological properties play a crucial role in preventing the annihilation of individual solitons or antisolitons. Defining the Chern 1-form, $\mathbf{A} = i A_\mu\mathrm{d}x^\mu$, is sufficient to calculate the topological property, the Chern number
    \begin{equation}\label{chernnumber}
        C_4 = \frac{1}{8\pi^2}\oint_{\partial M}\mathrm{Tr}\left(-\mathrm{d}\mathbf{A}\wedge\mathbf{A} + \frac{2}{3}\mathbf{A}\wedge\mathbf{A}\wedge\mathbf{A}\right)~,
    \end{equation}
on the surface $\partial M$ of a 4D region $M$. The minus sign in the first term comes from the minus sign in covariant derivative~(\ref{covariantderivative}). The covariant derivative (\ref{covariantderivative}) is invariant under a gauge transformation,
    \begin{equation}\label{gaugetransformation}
        \Phi\rightarrow \mathbf{g}\Phi~,~ A_\mu\rightarrow \mathbf{g}A_\mu\mathbf{g}^{-1} - i\partial\mathbf{g}\mathbf{g}^{-1}~.
    \end{equation}
On the surface $\partial M$ where the covariant derivative vanishes and the absolute value of the Higgs field is $1$ ($\phi_0\phi^0+\phi_a\phi^a=1$), the Chern number $C_4$ can only change by an integer through gauge transformation. This makes the variation of $C_4$ difficult.

\section{Spherical-symmetry solitons and antisolitons}
\label{sec3}
Due to spherical symmetry in 4D space, it is straightforward to set the ansatz for the Higgs field,
    \begin{equation}
        \label{solitonhiggsansatz}
        \Phi^\pm(r) = h(r) G_\pm\begin{pmatrix}1 \\ 0\end{pmatrix}~,~ G_\pm = \frac{1}{r}(i\tau_a x^a \pm x^4)~,
    \end{equation}
where $r=\sqrt{x_\mu x^\mu}$ is the distance from the origin and ``$\pm$'' correspond soliton and antisoliton respectively.  The ansatz for the gauge field is
    \begin{equation}
        \label{solitongaugeansatz}
        A_\mu^\pm (r) = -i[1 - \kappa(r)] \partial_\mu G_\pm G_\pm{}^{-1}~.
    \end{equation}
$\kappa$ and $h$ are the variation functionals and is unrelated to whether the solution is a soliton or an antisoliton. With the ansatz, action~(\ref{action}) of an $R$-radius $4$-ball is written as
    \begin{align}\nonumber
        S_E(R) =& 2\pi^2\int_0^R dr \bigg\{\frac{1}{2}r^3 h'^2 + \frac{3}{2}r h^2\kappa^2 \\
        \nonumber
        &+ \frac{1}{g^2}\left[\frac{24 (2 - \kappa)^2 (1 - \kappa)^2}{r} + 6 r\kappa'^2\right] \\
        \label{solitonaction}
        &+ \lambda r^3(h^2 - 1)^2\bigg\}
    \end{align}
where $\kappa'$ and $h'$ are the derivatives of $\kappa$ and $h$. $2\pi^2$ in action~(\ref{solitonaction}) comes from the surface area of the $r$-radius $4$-ball. Similar to the 2D and 3D cases, the distributions of $\kappa$ and $h$ that minimize action~(\ref{solitonaction}) also minimize the original action~(\ref{action}). 

\begin{figure}
    	\centering 
    	\includegraphics[width=0.4\textwidth, angle=0]{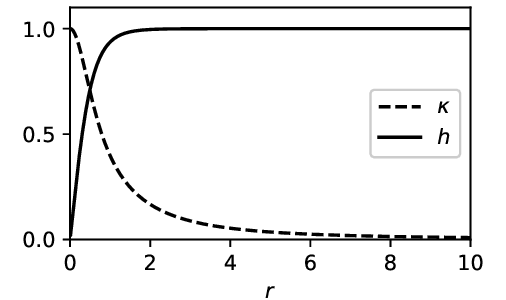}
    	\caption{The 4D soliton profile functions $h(r)$ and $k(r)$ for $\lambda=1,g=10$.} 
    	\label{numricalsoliton}%
    \end{figure}

The corresponding Euler-Lagrange equations of action~(\ref{solitonaction}) are
    \begin{subequations}\label{solitoneleqs}
    \begin{equation}
        r^3 h'' + 3r^2 h' - 3r h\kappa^2 - 4\lambda r^3 h (h^2 - 1) = 0, \label{higgseleq}
    \end{equation}
    \begin{equation}
        r\kappa'' + \kappa' - \frac{g^2 r\kappa h^2}{4} - \frac{8(1 - \kappa)(2 - \kappa)(2\kappa - 3)}{r} = 0. \label{kappaeleq}
    \end{equation}
    \end{subequations}
Setting $h$ and $\kappa$ on a lattice with $8000$ sites and lattice constant $\Delta r=0.01$, action~(\ref{solitonaction}) is minimized by gradient descent algorithm. The result is shown in Fig.~\ref{numricalsoliton}. The choices of $\lambda$ and $g$ are made to ensure that the ranges of variation for $\kappa$ and $h$ are comparable, enabling a better representation in Fig.~\ref{numricalsoliton}. While $r\rightarrow\infty$, $\kappa\sim 192/g^2 r^2$ and $h\sim 1-\exp(-2\sqrt{2\lambda}r)$. As a result, choosing an $R_0$ satisfying $R > R_0 \gg 1$, action~(\ref{solitonaction}) can be reduced to
    \begin{align}\nonumber
        S_E(R) =& \frac{192\pi^2}{g^2} \int_{R_0}^R dr \left[\frac{1}{r} + O\!\left(\frac{1}{r^3}\right)\right] + S_E(R_0) \\
        \label{largeRaction}
        =& \frac{192\pi^2}{g^2}\ln R + O\!\left(\frac{1}{R^2}\right) + S_0 ~.
    \end{align}
The absence of $R_0$ as a system parameter implies that the $R$-irrelevant term $S_0$ is not related to $R_0$. $S_E(R)$ diverges as $R$ tends to infinity, which means such a soliton or antisoliton cannot really exist in 4D flat space. The logarithmical divergence of action is similar to the energy divergence of global vortex \cite{PhysRevLett.48.1867}.

On the infinite 3-sphere $S_\infty^3$, $\kappa=0$ and $h=1$. The Chern number for the soliton/antisoliton is
    \begin{equation}\label{solitonchernnumber}
        C_4^\pm =\frac{1}{8\pi^2} \oint_{S_\infty^3}\!\!\!\frac{\pm 2\varepsilon_{ijkl}\,x^l\,\mathrm{d}x^i\!\wedge\!\mathrm{d}x^j\!\wedge\!\mathrm{d}x^k}{3r^4} = \pm 1~,
    \end{equation}
where $\varepsilon_{ijkl}$ is the Levi-Civita symbol.


\begin{table}
    \centering 
    \begin{tabular}{c c c c c c} 
        \hline
            $\lambda$ & $g$ & $192\pi^2/g^2$ &
            $E_0$ & $E_2$ & $4E_0+2E_2$\\
        \hline
            1   & 10 & 18.95 & 0.47 & 8.54  & 18.98 \\
            1   & 5  & 75.80 & 1.56 & 34.80 & 75.83 \\
            0.5 & 10 & 18.95 & 0.38 & 8.55  & 19.03 \\
            10  & 10 & 18.95 & 0.29 & 8.89  & 18.95 \\ 
        \hline
    \end{tabular}
    \caption{Single soliton's characteristic energies of scale transformation.}
    \label{single_numerical}
    \end{table}

The energy of a single soliton in the entire infinite space is infinite, which makes it challenging to analyze the scaling transformation. However, the soliton in a finite space, a $R$-radius 4-ball as previous, has finite energy. Now, we analyze the scaling transformation of the finite-space soliton. The setting of the spatial size $R$ renders the system inconsistent with Derrick's original scaling argument since the system changes with scaling parameter $\mu$. Following the convention in Ref.~\cite{manton}, the Euclidean action is decomposed into three parts: i) the Higgs potential part $E_0$, corresponding to the second term in Eq.~(\ref{action}); ii) the covariant momentum part $E_2$, corresponding to the first term in Eq.~(\ref{action}); and iii) the gauge part $E_4$, corresponding to the third term in Eq.~(\ref{action}). Due to Eq.~(\ref{largeRaction}), the gauge part $E_4$ can be understood as the integral from the radius $\xi$ to the boundary with radius $R$. $\xi$ is chosen to make $E_4$ satisfy
    \begin{equation}\label{singlegaugeseparation}
        E_4 = \frac{192\pi^2}{g^2}\ln \frac{R}{\xi},
    \end{equation}
and is proportional to the size of the soliton's core, where the gauge field is small. The core-related radius $\xi$ only depends on the absolute value of $\Phi$, which has been normalized to 1. With the scaling transformation $\mathbf{x}\rightarrow\mu\mathbf{x}$, the transformation of $\Phi$ is $\Phi\rightarrow\Phi$ and the single-soliton action transforms as
    \begin{equation}\label{singlescalingtransformation}
        S_E^p \rightarrow S_E^p(\mu) = \frac{1}{\mu^4}E_0 + \frac{1}{\mu^2}E_2 + \frac{192\pi^2}{g^2}\ln \frac{\mu R}{\xi}~,
    \end{equation}
since $R\rightarrow\mu R$ and $\xi\rightarrow\xi$. At the stationary point $\mu=1$, the requirement of $d S_E^p/d\mu=0$ leads to
    \begin{equation}\label{singleconstraint}
        4E_0 + 2E_2 = 192\pi^2/g^2~.
    \end{equation}
To check the constraint~(\ref{singleconstraint}), we numerically calculate the characteristic energies for given parameters $\lambda,g$ in Table~\ref{single_numerical}. Comparing the third and the sixth columns, Eq.~(\ref{singleconstraint}) is well satisfied. The single soliton in the $R$-radius 4-ball is an easily verifiable example where the energy, after scaling transformation, contains a $\ln \mu$ term in a system with a length characteristic.

\section{Soliton-antisoliton configuration}
\label{sec4}
When studying static interactions between particles and antiparticles \cite{manton, taubes1, taubes2, saurabh, chaves}, it is common to fix the particle and antiparticle at given points. The particle-antiparticle separation is a flexible parameter that can lower the total energy unless the derivative of the potential between the particle and the antiparticle is zero. This operation inevitably introduces some approximations. Such approximations are acceptable since we believe that perturbation theory is applicable in particle-antiparticle studies. From another view, a configuration that approximately satisfies the static Euler-Lagrange equations should have small time derivatives of the fields, when substituted into the time-dependent Euler-Lagrange equations. The small time derivatives imply that the configuration can maintain its shape for a considerable time. Therefore, a soliton-antisoliton configuration, that approximately satisfies the Euler-Lagrange equations of the action~(\ref{action}), is sufficient to study the static soliton-antisoliton interaction.

We define
    \begin{equation}\label{rho}
        \rho = \sqrt{x_a x^a}
    \end{equation}
and
    \begin{equation}\label{rprm}
        r_\pm = \sqrt{\rho^2 + (x_4 \mp s)^2} ~.
    \end{equation}
Assuming a soliton located at $(0,0,0,s)$ and an antisoliton at $(0,0,0,-s)$, a suitable configuration can have a finite Euclidean action. To describe the configuration of a soliton-antisoliton pair, we also need
    \begin{equation}\label{Gpair}
        G_p = \frac{r^2 - s^2 + 2is\tau_a x^a}{r_+ r_-}
    \end{equation}
like in Eqs.~(\ref{solitonhiggsansatz}) and (\ref{solitongaugeansatz}). Therefore, the ansatz of fields is
    \begin{subequations}\label{pairansatz}
    \begin{equation}\label{pairhiggsansatz}
        \Phi^p = h(\rho,x^4) G_p\begin{pmatrix}1 \\ 0\end{pmatrix}~,
    \end{equation}
    \begin{equation}\label{pairgaugeansatz}
        A_\mu^p = -i[1 - \kappa(\rho,x^4)] \partial_\mu G_p G_p{}^{-1}~.
    \end{equation}
    \end{subequations}
Ansatz~(\ref{pairansatz}) has rotation symmetry around $x_4$ axis. By substituting ansatz~(\ref{pairansatz}) into action~(\ref{action}), we obtain
    \begin{align}\nonumber
        S_E^p =& 4\pi\!\int\! \rho^2 d\rho dx_4\Bigg\{\frac{1}{2}\!\left(\frac{\partial h}{\partial\rho}\right)^2 + \frac{1}{2}\!\left(\frac{\partial h}{\partial x_4}\right)^2
                + \frac{6 s^2 h^2 \kappa^2}{r_+^2 r_-^2} \\
        &+ \lambda (h^2 - 1)^2 + \frac{384s^4 (\kappa-1)^2 (\kappa-2)^2}{g^2 r_+^4 r_-^4} \nonumber \\
        &+ \frac{8s^2}{g^2 r_+^4 r_-^4}\Bigg[3\!\left((s^2 + \rho^2 - x_4^2)\frac{\partial\kappa}{\partial x_4} - 2\rho x_4 \frac{\partial\kappa}{\partial\rho}\right)^2 \nonumber \\
        &+ 2\!\left(2\rho x_4\frac{\partial\kappa}{\partial x_4} + (s^2 + \rho^2 - x_4^2)\frac{\partial\kappa}{\partial\rho}\right)^2 \Bigg] \Bigg\}~.
        \label{pairaction}
    \end{align}
$4\pi$ comes from the surface area of the $\rho$-radius $3$-ball. 

We investigate the well-separated situation by assuming the soliton-antisoliton pair's separation $2s$ is large enough. The entire 4D space can be divided into 3 parts: (i) a 4-ball with radius $R$ centered at $(0,0,0,s)$, symbolized as $\mathcal{B}_+^4$; (ii) a 4-ball with radius $R$ centered at $(0,0,0,-s)$, symbolized as $\mathcal{B}_-^4$; and (iii) the rest of the space, symbolized as $\mathcal{M}_r$. $R$ is chosen to satisfy $s\gg R\gg 1$. 

In $\mathcal{B}_+^4$, $r_+<R\ll s$ and $r_-\approx 2s$. With $\varepsilon=x_4-s \ll s$, Eq.~(\ref{Gpair}) is approximately as
    \begin{align}
        G_p \approx & \frac{x_a x^a + (s + \varepsilon)^2 - s^2 + 2is\tau_a x^a}{2sr_+} \nonumber \\
        \label{approxGp}
        \approx & \frac{i\tau_a x^a + \varepsilon}{r_+}~,
    \end{align}
which has the same form of $G_+$ in Eq.~(\ref{solitonhiggsansatz}). In Eq.~(\ref{approxGp}), terms of order $\varepsilon/s$, $r_+/s$, and of higher orders have been neglected. In $\mathcal{B}_-^4$, the situation is similar and
    \begin{equation}\label{approxGm}
        G_p \approx \frac{i\tau_a x^a - \varepsilon}{r_-}
    \end{equation}
with $\varepsilon=x_4+s$. Therefore, fields~(\ref{pairansatz}) in $\mathcal{B}_\pm^4$ are approximately the same fields in Eqs.~(\ref{solitonhiggsansatz}) and (\ref{solitongaugeansatz}). This means that the ansatz~(\ref{pairansatz}) describes a soliton/antisoliton with accuracy of $R/s$ near $(0,0,0,\pm s)$. The integral~(\ref{pairaction}) in $\mathcal{B}_\pm^4$ is
    \begin{equation}\label{ballintegral}
        S_E^p\left(\mathcal{B}_\pm^4\right) \approx \frac{192\pi^2}{g^2}\ln R + S_0 ~.
    \end{equation}
The center of the soliton/antisoliton is fixed at $(0, 0, 0, \pm s)$ by the determinacy of the fields at these points.

In $\mathcal{M}_r$, $h\approx 1$ and the dominant term of $\kappa$ is
    \begin{equation}\label{kappaapprox}
        \kappa \approx \frac{384 s^2}{g^2 r_+^2 r_-^2} \leq \frac{96}{g^2} \left(\frac{1}{r_+} + \frac{1}{r_-}\right)^2 \ll 1~.
    \end{equation}
Considering $x_\mu/s$ and $R/s$ as new variables in the integral~(\ref{pairaction}), the dominant term is
    \begin{align}\nonumber
        S_E^p\left(\mathcal{M}_r\right) \approx& 4\pi\int_{\mathcal{M}_r} d^4 r ~ \frac{384s^4 \cdot 4}{g^2 r_+^4 r_-^4} \\
        \label{restintegral}
        =& \frac{384\pi^2}{g^2} f(R/s)~.
    \end{align}
Since $x_\mu/s$ has already been integrated out, $f$ only depends on $R/s$. The specific expression of $f$ will be given after Eq.~(\ref{premainpairaction}). While $r_\pm\gg s$, $G_p$ is close to $1$ and the fields~(\ref{pairansatz}) are approximately the vacuum fields. This implies that the nonzero action is predominantly concentrated within a finite region of space. Therefore, the total action is finite.

For the ansatz~(\ref{pairansatz}), the total Chern number is
    \begin{equation}\label{pairchernnumber}
        C_4^p = 0~,
    \end{equation}
since the fields at infinity correspond to the trivial vacuum state. However, according to Eqs.~(\ref{approxGp}) and (\ref{approxGm}), the Chern number approaches $\pm 1$ for an $R$-radius 3-sphere ($1\ll R<s$) near $(0, 0, 0, \pm s)$. 

Combining Eqs.~(\ref{ballintegral}) and (\ref{restintegral}), the main parts of action~(\ref{pairaction}) is
    \begin{equation}\label{premainpairaction}
        S_E^p \approx \frac{384\pi^2}{g^2}\left[f(R/s) + \ln R \right] + 2S_0 ~.
    \end{equation}
Since the choice of $R$ is arbitrary, $S_E^p$ should be independent of $R$ and $f$ should be $f(x)=-\ln x + C$. The final result is
    \begin{equation}\label{mainpairaction}
        S_E^p \approx \frac{384\pi^2}{g^2}(\ln s + C) + 2S_0 ~.
    \end{equation}
This means that well-separated soliton and antisoliton exhibit an dominant attractive interaction potential that is logarithmically dependent on their separation. Compared to the linear growth of the quark-antiquark potential with separation, the action in Eq.~(\ref{mainpairaction}) increases logarithmically. Although soliton-antisolition potential does not diverge as rapidly as the quark-antiquark one, it still diverges as the separation approaches infinity. This means that soliton-antisoliton pairs exhibit a weak confinement. Free solitons cannot exist independently, and soliton-antisoliton pairs can exist. The logarithmical interaction here is analogous to the global vortex-antivotex interaction in the Goldstone model \cite{PhysRevD.47.1324,Ovchinnikov_1998}.

To minimize action~(\ref{pairaction}), $h$ and $\kappa$ are set on a lattice with $768\times 128$ sites and lattice constants $\Delta x_4=\Delta\rho=0.1$. On the lattice, $x_4$ changes from $-38.35$ to $38.35$ and $r$ changes from $0$ to $12.7$. The separation is chosen to satisfy $12.5<s<25.5$ to ensure they are neither close to the boundary nor close to each other. By choosing a larger value of $g$, the total action can exhibit form~(\ref{mainpairaction}) over a shorter separation, allowing numerical computations to achieve higher accuracy near the center of the soliton/antisoliton. We minimize action~(\ref{pairaction}) by gradient descent algorithm with open boundary conditions. We note that the contribution from the outside area of the lattice is also important to obtain the total action, however, such contribution can be easily given by Eq.~(\ref{restintegral}). 

\begin{figure}
    	\centering 
    	\includegraphics[width=0.4\textwidth, angle=0]{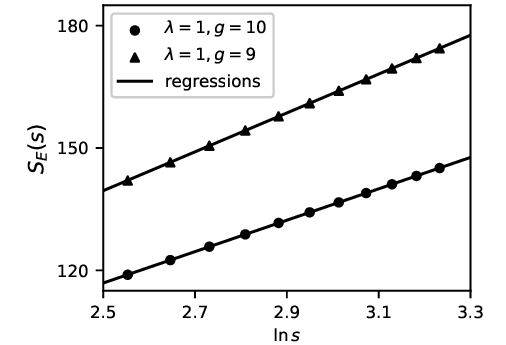}
    	\caption{The total action $S_E^p$ as a function of the soliton-antisoliton’s separation $s$.} 
    	\label{regressions}%
    \end{figure}

The linear regressions of the total action $S_E^p$ with respect to $\ln s$ are shown in Fig.~\ref{regressions}. There are only two sets of parameters are presented in Fig.~\ref{regressions} for clarity. More linear regression slopes and intercepts are listed in the fourth and fifth columns of Table~\ref{pair_numerical}. It is easy to observe that the slopes match the expectations of Eq.~(\ref{mainpairaction}) within the error of $1/s$.

    \begin{table*}
        \centering 
        \begin{tabular}{cccccccc}
            \hline
            $\lambda$ & $g$ & $384\pi^2/g^2$ & \textrm{Slope} & \textrm{Intercept} &
            $E_0$ & $E_2$ & $4E_0+2E_2$\\
            \hline
            1   & 10 & 37.90  & 38.50  & 20.63  & 0.94 & 17.22 & 38.20 \\
            1   & 9  & 46.79  & 47.66  & 20.37  & 1.14 & 21.31 & 47.18 \\
            1   & 5  & 151.60 & 158.06 & -30.61 & 3.09 & 70.46 & 153.28 \\
            0.1 & 10 & 37.90  & 38.51  & 18.61  & 0.70 & 17.69 & 38.18 \\
            10  & 10 & 37.90  & 38.50  & 22.47  & 0.59 & 17.91 & 38.18 \\
            \hline
        \end{tabular}
    \caption{Parameters of linear regression and the characteristic energies of scale transformation.}
    \label{pair_numerical}
    \end{table*}

Since the approximations in ansatz~(\ref{pairansatz}) does not hold when soliton and antisoliton are close, we cannot determine the nature of their interaction at close separations. Therefore, the existence of the static saddle point particle is uncertain.

Next, we discuss the scaling transformation of the soliton-antisoliton pairs. The setting of the separation distance $s$ challenges the assumption of traditional scaling argument because $s$ is different after a scaling transformation. Like in Sec.~\ref{sec3}, the Euclidean action with ansatz~(\ref{pairansatz}) is decomposed into $E_0$, $E_2$ and $E_4$. While $s$ is large enough, $E_0$ and $E_2$ are almost independent on $s$ and their average values are listed in the sixth and seventh columns of Table~\ref{pair_numerical}. Due to Eqs.~(\ref{largeRaction}), (\ref{restintegral}), and (\ref{mainpairaction}), 
    \begin{equation}\label{gaugeseparation}
        E_4 = \frac{384\pi^2}{g^2}\ln s + E_4^\prime~,
    \end{equation}
where $E_4^\prime$ does not depend on $s$. $E_4$ can be understood as the integral of the gauge field action density, excluding the soliton and antisoliton core regions. This integral is linear to $\ln s/\xi$. With the scaling transformation $\mathbf{x}\rightarrow\mu\mathbf{x}$, the soliton-antisoliton action transforms as
    \begin{equation}\label{scalingtransformation}
        S_E^p \rightarrow S_E^p(\mu) = \frac{1}{\mu^4}E_0 + \frac{1}{\mu^2}E_2 + \frac{384\pi^2}{g^2}\ln \mu s + E_4^\prime~,
    \end{equation}
since $s\rightarrow\mu s$. The stationary constraint $d S_E^p/d\mu=0$ at $\mu=1$ leads to
    \begin{equation}\label{constraint}
        4E_0 + 2E_2 = 384\pi^2/g^2~.
    \end{equation}
This is numerically checked by the last column in Table~\ref{pair_numerical} with error of $1/s$. As discussed in Ref.~\cite{manton}, the lack of $\ln\mu$ term leads to the action's instability under scaling transformation. In a system where the soliton-antisoliton separation is given, the existence of such a term allows for a meaningful discussion of four-dimensional solitons with an adjoint Higgs field. For suitable state in an appropriate 4D Higgs-type model, such a term can exist, which is essential for the discussions of 4D solitons with an adjoint Higgs field.

\section{Conclusions}
\label{sec5}
For the model described by the action~(\ref{action}), we have investigated the static solutions with spherical symmetry. These solutions have Chern numbers $\pm 1$ and correspond to a soliton and an antisoliton. In the entire infinite space, the action of such solutions diverges, indicating that these solutions cannot truly exist. However, in a finite space, such as an R-radius 4-ball, these solutions have finite action. We have also analyzed the scaling transformation of the finite-space single-soliton state and obtained the stability condition, which is also verified numerically.

When a soliton and an antisoliton coexist, they can form a quasi-static state with finite action. Such a state is topologically trivial, as its total Chern number is 0. The action is linearly proportional to the logarithm of the separation. We have verified this interaction numerically. As the separation tends to infinity, the action diverges, reiterating that a soliton/antisoliton cannot exist independently. We have also verified the scaling stability condition of the soliton-antisoliton pair.

\section*{Acknowledgements}
This work is supported by the National Key R\&D Program of China (Grant No. 2022YFA1403601). Liang Gong thanks Prof. Paul Sutcliffe at Durham University for discussions.

\bibliographystyle{elsarticle-num} 
\bibliography{WSSAPAHF}






\end{document}